\documentclass[prl,twocolumn,showpacs,floatfix,amsmath,amsfonts]{revtex4}

\usepackage{amsmath}
\usepackage{epsfig}

\begin{document}

\title{Soliton molecules in trapped vector Nonlinear Schr\"odinger systems}


\author{V\'{\i}ctor M. \surname{P\'erez-Garc\'{\i}a}}
\affiliation{Departamento de Matem\'aticas, Escuela T\'ecnica
Superior de
  Ingenieros Industriales, \\
  Universidad de Castilla-La Mancha 13071 Ciudad Real, Spain}
\author{Vadym \surname{Vekslerchik}}
\affiliation{Departamento de Matem\'aticas, Escuela T\'ecnica
Superior de Ingenieros Industriales, \\
  Universidad de Castilla-La Mancha 13071 Ciudad Real, Spain}
  \altaffiliation[Permanent address:]{Institute for Radiophysics and
  Electronics, Kharkov 61085, Ukraine}
\date{\today}


\begin{abstract}
We study a new class of vector solitons in trapped Nonlinear
Schr\"odinger systems modelling the dynamics of coupled light
beams in GRIN Kerr media and atomic mixtures in Bose-Einstein
condensates. These solitons exist for different spatial
dimensions, their existence is studied by means of a systematic
mathematical technique and the analysis is made for inhomogeneous
media.
\end{abstract}

%

\maketitle


 Since the begining of its history, physics
has studied simple objects and the way they arrange to form more
complex structures. Some remarkable successes included the atomic
theory of matter, the structure of nucleus in terms of protons and
neutrons and the substructure of nucleons in terms of quarks to
cite a few examples.

Elementary {\em robust} objects made of light have been known
since the 70's. In fact, {\em spatial optical solitons}---
self-trapped states of light with particle-like properties--- have
attracted a considerable attention during last years as possible
building blocks of all-optical switching devices where light is
used to guide and manipulate light itself \cite{steg,science}.
Another field where robust solitonic structures have been recently
found is that of Bose-Einstein condensation where the dilute
quantum gas supports robust structures such as one-dimensional
dark solitons \cite{dark}, bright solitons \cite{bright} or
vortices \cite{Fetter}.

In nonlinear optics, the robust nature of spatial optical solitons
\cite{science}, allows to draw an analogy with atomic physics
treating spatial solitons as ``atoms of light''. Furthermore, when
several light beams are combined to produce {\em a vector
soliton}, this process can be viewed as the formation of composite
states or ``molecules of light''.

Several structures of this type have been studied previously:
dipole and multipole vector solitons
\cite{us,exper,multipole,moti}, self-trapped necklace ring beams
\cite{Segev1}, rotating propeller solitons \cite{Carmon} and
rotating optical soliton clusters \cite{Yuri3}.

In the field of Bose-Einstein condensation (BEC) vortex solitons
arrange to generate lattice structures in rotating \cite{ENS} and
non-rotating scalar systems \cite{Torner}. In multicomponent
condensates simple stationary solutions involving some type of
dynamical equilibria between the constituent solitons have been
described \cite{PRLdual,Nature,Malomed}.

The purpose of this letter is to describe and analyze in detail a
way to build multidimensional ``soliton molecules" in vector
systems with applications to nonlinear optics and to Bose-Einstein
condensed gases. As new features these solitons will be analyzed
using a systematic mathematical technique, the analysis will be
made for inhomogeneous media and the soliton molecules exist for
different dimensions $D = 2,3$ thus providing the first soliton
molecules shown to exist for $D=3$.

{\em The model.-} We will consider a system of $N$ complex fields
$u_{1}(t,\mathbf{r}), u_2(t,\mathbf{r}), ..., u_{N}(t,\mathbf{r})$
ruled by the equations
\begin{equation}
  i \partial_{t} \; u_{j}(t,\mathbf{r}) =
  \left[
    - {1 \over 2} \Delta + V(\mathbf{r}) + U_{j}(t,\mathbf{r})
  \right]
  u_{j}(t,\mathbf{r}), \label{basic}
\end{equation}
for $j = 1,...,N$. The coupling term is given by $
U_{j}(t,\mathbf{r}) =
  \sum_{k} g_{jk}
    \left|u_{k}(t,\mathbf{r}) \right|^{2}$ with $g_{jk} \in
    \mathbb{R}$.
Eqs. (\ref{basic}) are a set of Nonlinear Schr\"odinger equations
(NSE) which in BEC problems describe multicomponent systems,
$u_{j}$ being the wavefunctions for each of the atomic species
involved \cite{PRLdual,Nature,Malomed}. In optics these equations
describe the incoherent interaction between the slowly varying
envelopes of the electric field in paraxial beams in Kerr media.
We choose $V(\mathbf{r}) = \mathbf{r}^2/2$ which corresponds to an
isotropic magnetic trapping in BEC and to a GRIN fiber in the
optical case.

\noindent {\em Single component case: Soliton atoms.-}  Let us
first consider the scalar case ($N=1$). Solitons or stationary
solutions of Eq. (\ref{basic}) in the scalar case have the form
$u(t,\mathbf{r}) = \phi_{\mu}(\mathbf{r}) e^{i \mu t}$ and satisfy
\begin{equation}
\label{single} \mu \phi_{\mu} = -\frac{1}{2}\Delta \phi_{\mu} +
\frac{1}{2}\mathbf{r}^2 \phi_{\mu} + g |\phi_{\mu}|^2 \phi_{\mu}
\end{equation} For a fixed norm of the solution $\|\phi\|_{L^2} :=
\int |\phi|^2 dV$, any solution to Eq. (\ref{single}) will be a
valid soliton atom for our purposes. The simplest case corresponds
to a nodeless ground state solution which for strong interaction
is close to the Thomas-Fermi solution (thus a quasi-compacton type
of solution) \cite{Malomed} and in the small interaction case (as
it happens in nonlinear optics) it is close to a Gaussian
function. Many other complex stationary solutions to Eq.
(\ref{single}) are possible such as vortices. These objects will
play the role of ``atoms" in what follows.

The relevant property to be used here  \cite{VJV}, is that any
function of the form
\begin{equation}
u({\bf r},t) = \phi_{\mu}\left({\bf r}-{\bf R}(t)\right)
e^{i\left[\mu t + \theta({\bf r},t)\right]}, \label{traslada}
\end{equation}
is a solution of the scalar time-dependent NLSE provided
$\mathbf{R}(t)$ satisfies $\tfrac{d^2}{dt^2}\mathbf{R}  +
\mathbf{R} = 0$ and $\theta(\mathbf{r},t) =
(\mathbf{r},\tfrac{d}{dt}\mathbf{R}) + f(t)$, where $f(t) =
\int_0^t \left[\left(\tfrac{d}{dt}\mathbf{R}\right)^2 -
\mathbf{R}^2 \right]dt$. This means that exact scalar
time-dependent solutions whose center evolves according to
harmonic oscillator type equations and {\em preserve the shape of
the stationary solution} during evolution can be built.
Remarkably, this property is not exclusive of the Kerr
nonlinearity and it is valid for any type of nonlinear term for
which localized solutions exist.
   However, if $N\geq2$,
  the property described previously does not
hold except for very trivial situations. Because of the
cross-interaction. In a general situation, the pulses collide and
lose their individuality. Here we want to give some ideas on how
to overcome this problem and build stationary nontrivial vector
solitons of Eqs. (\ref{basic}).

\noindent {\em Formalism for the multicomponent case.-} Let us
define the modulus, $n_j$, and phase, $\phi_j$, of each wavepacket
through $u_{j} = \sqrt{n_{j}} \exp\left( i\varphi_{j} \right)$.
Let us define also the center of mass of species $j$,
$\mathbf{R}_{j}(t) = \int dV \; n_{j} \, \mathbf{r}$ and their
total momenta $\mathbf{P}_{j}(t) = \int dV \; n_{j} \nabla
\varphi_{j}$, whose evolution laws are
\begin{eqnarray}
  \frac{d}{dt} \, \mathbf{R}_{j} & = & \mathbf{P}_{j},
\\
\label{dPj}
  \frac{d}{dt} \, \mathbf{P}_{j} & = &
  - \mathbf{R}_{j}
  + {1 \over 2} \sum_{k} g_{jk} \mathbf{F}_{jk}
\end{eqnarray}
The first term in the r.h.s of (\ref{dPj}) corresponds to the
external potential, while the nonlinear force is given by
\begin{equation}
  \mathbf{F}_{jk} =
  \int dV \; \left( n_{k} \nabla n_{j} - n_{j} \nabla n_{k} \right)
\label{force}
\end{equation}
Generally $\mathbf{F}_{jk}$ may be a rather complex function, but
if the $n_j$ for $j = 1,...,N$, have small enough overlapping, it
is possible to argue that these forces would be central. To prove
this affirmation let us first notice that for the scalar case and
far from the center of the wavepacket, the self-interaction is
small and $u$ can be described by the linear theory to be $ n
\left( \mathbf{r}\right) \propto
  e^{- r^{2}} \left[r^{2m}
  + \mathcal{O}\left(r^{2m-2}\right)\right]$ for $r \to \infty$.
Similar considerations apply to the multicomponent case. If the
wavepackets are separated, then in the region where the $u_j$
overlap (i.e., the region which determines the value of the
integral in (\ref{force})) the nonlinear terms will be small and
\begin{equation}
  \nabla n_{j} = -2\left(\mathbf{r} - \mathbf{R}_{j}\right)
  + \mathcal{O}\left(\frac{1}{\left|\mathbf{r} -
  \mathbf{R}_{j}\right|}\right),
\end{equation}
from which it follows that
\begin{equation}
  \mathbf{F}_{jk} =
  2\left(\mathbf{R}_{j} - \mathbf{R}_{k}\right)
  \int dV \; n_{j} n_{k}
  + \mathcal{O}\left(\frac{1}{\left|\mathbf{R}_{k} -
  \mathbf{R}_{k}\right|}\right),
\end{equation}
i.e., if the wavepackets are separated, in the leading order
\emph{the inter-mode force is central}.
\begin{figure}
\epsfig{file=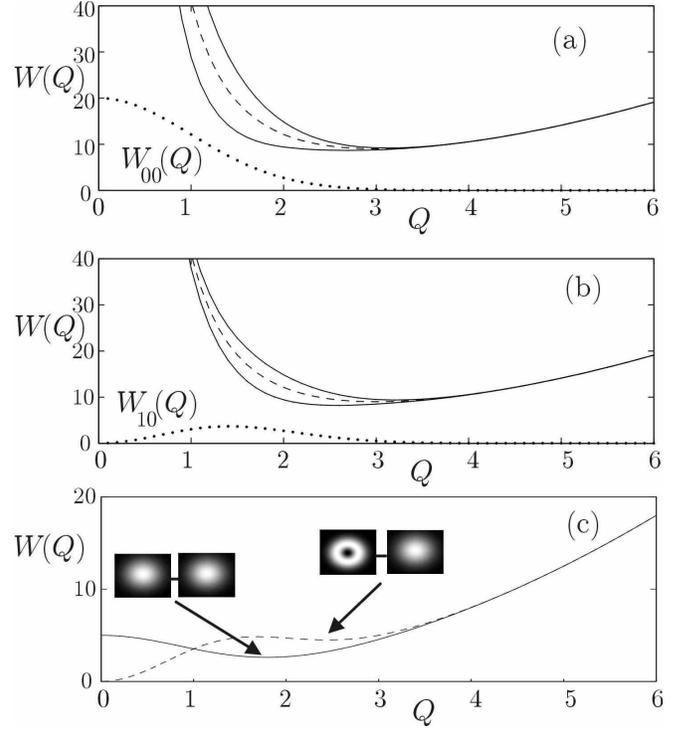,width=\columnwidth} \caption{Potential
$W(Q)$ for $g_{jk}=0$ (dashed line) and nonlinear potential
$W_f(Q)$ (dotted line) for: (a) Interaction of two gaussian modes,
(b) Interaction of a gaussian with a vortex mode. Solid lines
represent the total effective potential for $\bar{g}/2\pi =  20$
(upper solid line) or $\bar{g}/2\pi = -20$ (lower solid line) (c)
A situation with $L=0$. Solid line: total potential for two
gaussian modes with $\bar{g}/2\pi =  5$. Dashed line: potential
for gaussian plus vortex modes with $\bar{g}/2\pi =  20$. The
arrows mark the minima corresponding to non-rotating vector
molecules\label{prima}}
\end{figure}

As to the factor $\int dV \; n_{j} \, n_{k}$, let us evaluate it,
for illustrative purposes, for the case when $u_{j,k}$ are
solutions of linear problem ($g_{jk}=0$) of the type $  n_{j}  =
    N_{j}
    \left| \mathbf{r} - \mathbf{R}_{j} \right|^{2m_{j}}
   e^{- \left( \mathbf{r} - \mathbf{R}_{j} \right)^{2}}$, where $N_{j}$
   are normalization constants (these functions
    include the fundamental mode $m_j=0$ and vortex states)
\begin{equation}
\label{forces}
  \int dV \; n_{j} n_{k} =
  \frac{\pi}{2}
  N_{j} N_{k}
  e^{-\tfrac{1}{2}\left| \mathbf{R}_{j} - \mathbf{R}_{k}
  \right|^{2}}
  K_{ m_{j}, m_{k} } (\left| \mathbf{R}_{j} - \mathbf{R}_{k}
  \right|),
\end{equation}
where $K_{ m_{j}, m_{k} }$ are polynomial factors, the lowest
order ones for $D=2$ being $ K_{ 0, 0 } (Q) = 1$, $K_{ 0, 1 } (Q)
= Q^{2}/4$ and $K_{ 1, 1 } (Q) =  \left[
    2 -  Q^{2} + \tfrac{1}{4} Q^{4} \right]/4$.
Thus, in the case of small nonlinearities, we can estimate the
force between the wavepackets by using Eq. (\ref{forces}). For our
purposes the specific form of the interaction is less crucial than
the fact that the forces $\mathbf{F}_{jk}$ are central.

{\em Two-component case.-}
  For the two-component symmetric ($g_{12}=g_{21}=\bar{g}$) case Eqs. (\ref{dPj})
  read
\begin{subequations}
\begin{eqnarray}
  \frac{d}{dt} \, \mathbf{P}_{1} & = &
  -  \, \mathbf{R}_{1}
  + {1 \over 2} \bar{g} \mathbf{F},
\\
  \frac{d}{dt} \, \mathbf{P}_{2} & = &
  -  \, \mathbf{R}_{2}
  - {1 \over 2} \bar{g} \mathbf{F},
\end{eqnarray}
\end{subequations}
where $ \mathbf{F} =
  \int dV \; \left( n_{2} \nabla n_{1} - n_{1} \nabla n_{2}
  \right)$.
The modified 'total' center of mass   $\mathbf{R}  =
\mathbf{R}_{1} + \mathbf{R}_{2}$ and momentum $\mathbf{P}  =
\mathbf{P}_{1} + \mathbf{P}_{2}$ do not 'feel' the force
$\mathbf{F}$ since they satisfy $\tfrac{d}{dt}\mathbf{R} =
\mathbf{P}$ and $\tfrac{d}{dt} \mathbf{P} = - \mathbf{R}$, i.e.
they oscillate harmonically.

The most interesting quantity is $\mathbf{Q} = \mathbf{R}_{1} -
\mathbf{R}_{2}$ (for the general multicomponent case we define
$\mathbf{Q}_{ij}(t)=\mathbf{R}_i(t)-\mathbf{R}_j(t)$), which gives
the separation between the centers of mass of the two components
and evolves according to
\begin{equation}
\frac{d^2\mathbf{Q}}{dt^2} + \mathbf{Q} = \bar{g} \mathbf{F}
\label{main-eq}
\end{equation}
\begin{figure}
\epsfig{file=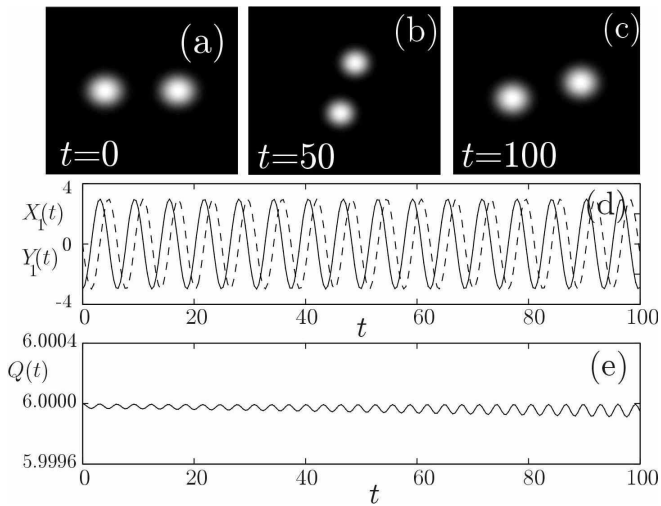,width=\columnwidth} \caption{Evolution
of initial data $u_1(\mathbf{r},t) =
\phi_0(\mathbf{r}-\mathbf{R}_1(0))e^{3iy}$, $u_2(\mathbf{r},t) =
\phi_0(\mathbf{r}+\mathbf{R}_1(0))e^{-3iy}$ with $g_{11} = g_{12}
= g_{22} = 10$, $\int n_j dV = 1$, $\mathbf{R}_1(0) = (3,0)$ and
$\phi_0$ is the scalar ground state. (a-c) Density plots of
$n_1(\mathbf{r},t)+n_2(\mathbf{r},t)$ on the spatial region
$[-8,8]\times[-8,8]$. (d) Evolution of $\mathbf{R}_1(t)$: $X_1(t)$
(solid) and $Y_1(t)$ (dashed). (e) Evolution of $Q(t)$.
\label{dual}}
\end{figure}
As discussed above, if $\mathbf{Q}$ is sufficiently large, the
force $\mathbf{F} \propto \mathbf{Q}$ and it can be presented in
potential form $\mathbf{F} =
  - \tfrac{\partial}{\partial \mathbf{Q}} W_{\mathrm{f}}(Q) =
  - W_{\mathrm{f}}'(Q) \mathbf{Q}/Q$
and then Eq. (\ref{main-eq}) reads
\begin{equation}
  \frac{d^2\mathbf{Q}}{dt^2} +
  W_{\mathrm{tot}}'(Q) \frac{\mathbf{Q}}{Q} = 0
  \label{effective}
\end{equation}
where $W_{\mathrm{tot}} =  \tfrac{1}{2}Q^{2} +
  \bar{g} W_{\mathrm{f}}(Q)$. It is interesting to note that in the approximation of
independent wavepackets
 the potential $W_{\mathrm{f}}$ is given by $
W_{\mathrm{f}}(Q) = \int dV \; n_{1} n_{2}$.

  It is evident that Eq. (\ref{effective}) has two
constants of motion: energy $E =
  \tfrac{1}{2}(\dot{\mathbf{Q}}, \, \dot{\mathbf{Q}}) +
  W_{\mathrm{tot}}(Q)$ and angular momentum, $ \mathbf{L} = \mathbf{Q}
  \times \dot{\mathbf{Q}}$. Since $
  \mathbf{\dot{Q}}^2 = \dot{Q}^{2} + L^{2}/Q^{2}$,
we can write $E =
  \tfrac{1}{2} \dot{Q}^{2} +
  W_{\mathrm{eff}}(Q)$, where
$
  W_{\mathrm{eff}}(Q) = \frac{1}{2}Q^{2} +
  { L^{2} \over 2Q^{2} } +
  \bar{g} W_{\mathrm{f}}(Q)$.
 \begin{figure}
\epsfig{file=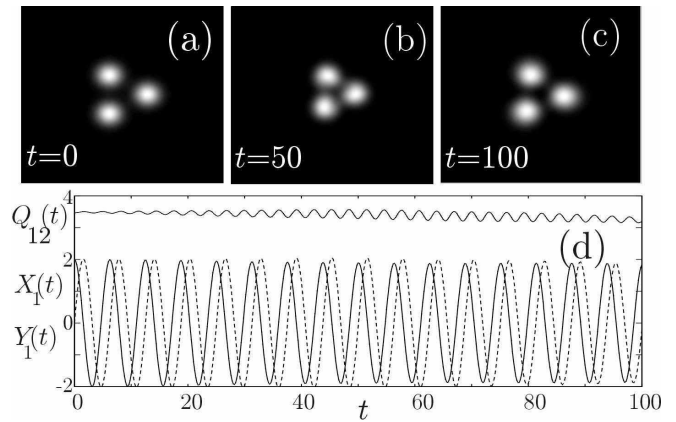,width=\columnwidth} \caption{Evolution
of a three-component system with $u_j(\mathbf{r},0) =
\left(1/\sqrt{\pi}\right)
e^{-((x-x_j)^2+(y-y_j)^2)/2}e^{i(v_{xj}x+v_{yj}y)}$, $(x_j,y_j) =
d_0(\cos(2\pi j/3),\sin(2\pi j/3))$, $(v_{xj},v_{yj}) = d_0(\sin
(2\pi j/3),-\cos(2\pi j/3))$ for $j = 0,1,2$. Parameter values:
$g_{ij} = 3, d_0 = 2, \int n_j dV = 1$. (a-c) Density plots of
$\sum_{j=1}^3 n_j(\mathbf{r},t)$ (d) Evolution of $\mathbf{R}_1$:
($X_1(t),Y_1(t)$) and $Q_{12}(t)$. \label{tertia}}
\end{figure}
There will exist an equilibrium distance $Q_0$, for which the
effective potential $W_{\mathrm{eff}}$ is minimized. The reason is
that the centrifugal contribution is divergent for $Q\simeq 0$
while the trap potential is unbounded for $Q \rightarrow \infty$
and the effective nonlinear interaction should
 decay for large $Q$ and have a maximum finite amplitude [Fig. \ref{prima}].

 In particular, if the nonlinear interaction is small then
$Q_0^{\text{lin}} \simeq L^{1/2}$
 thus leading to a stationary rotating solution of the vector system
(\ref{basic}) provided the distances between the components are
kept large enough. The larger the nonlinear terms the larger will
the deviation of $Q_0$ from $Q_0^{\text{lin}}$ be ($Q_{eq}
> Q_0$ for $g_{jk}>0$ and $Q_{eq}<Q_0$ for $g_{jk}<0$) [Fig
\ref{prima}(a,b)]. When $L=0$, the combination of the trap force
and the nonlinear term may also have minima [Fig. \ref{prima}(c)]
 corresponding to non-rotating soliton molecules. However,
due to the fact that the equilibrium distance is comparable to the
wavepacket's widths, our result must be taken only as an
indication that such soliton molecules exist and not as a proof
that they will be well approximated by combinations of scalar
solitons.
\begin{figure}
\epsfig{file=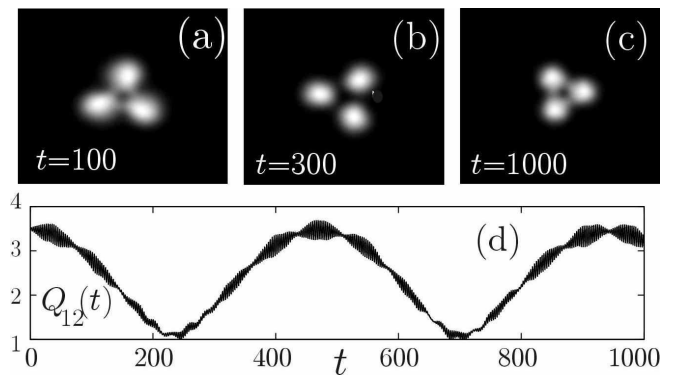,width=\columnwidth} \caption{Same as
Fig. \ref{tertia} but with $g_{ij} = 5$. (a-c) Density plots of
$\sum_{j=1}^3 n_j(\mathbf{r},t)$. (d) Long-time evolution of
$Q_{12}(t)$. \label{cuarta}}
\end{figure}

\noindent {\em Examples of soliton molecules.-} Let us now present
several examples of the soliton molecules discussed previously.
First we have studied the case of a pair of weakly interacting
soliton atoms. The results, obtained with a symplectic second
order in time split-step integrator are summarized in Fig.
\ref{dual} where it is seen how the small interaction induces only
small harmonic oscillations of $Q(t)$ without appreciable
distortion of the wavepackets.

\begin{figure}
\epsfig{file=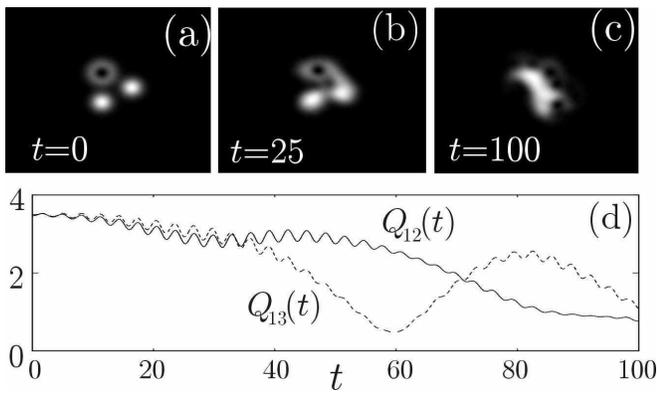,width=\columnwidth} \caption{Same as
Fig. \ref{cuarta} but with a gaussian soliton replaced by a vortex
soliton. (a-c) Density plots of $\sum_{j=1}^3 n_j(\mathbf{r},t)$.
(d) Evolution of the inter-mode distances $Q_{12}(t)$ (solid), and
$Q_{13}(t)$ (dashed).
 \label{quinta}}
\end{figure}
\begin{figure}
\epsfig{file=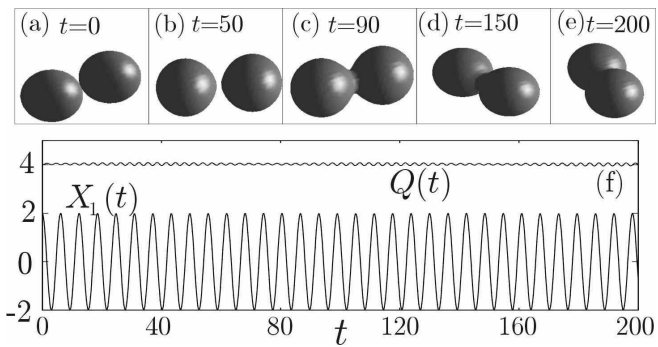,width=\columnwidth} \caption{Weak
interaction of two gaussian solitons $u_1(\mathbf{r},0) =
(1/\pi^{3/4})e^{-((x-2)^2+y^2+z^2)/2}e^{-2iy}$, $u_2(\mathbf{r},0)
= (1/\pi^{3/4})e^{-((x+2)^2+y^2+z^2)/2}e^{2iy}$ with $g_{jk} =
10$. (a-e) Isosurface plots of $n_1(\mathbf{r},t) +
n_2(\mathbf{r},t)$ for $n_1+n_2 = 0.01$. (f) Evolution of the
inter-mode distance $Q(t)$ (upper solid line), and $X_1(t)$ (lower
solid line).
 \label{sixta}}
\end{figure}

In Fig. \ref{tertia} three gaussian solitons interact more
strongly due to the larger number of components and the smaller
distance between the beams. In this case
 the oscillations of the distances between components, $Q_{ij}$, remain small
[Fig. \ref{tertia}(d)] although some oscillation of the positions
of the beams is appreciable [Fig. \ref{tertia}(a-c)]. Although
this particular configuration is stable, the present example is a
three-body problem for which many behaviors are possible: stable
solutions, resonances, chaos, etc. In fact, if the values of the
nonlinear coefficient are increased the beams deform and the
inter-mode distances $Q(t)$ suffer strong oscillations [Fig.
\ref{cuarta}(d)] although the structure remains stable after long
periods of time containing about one hundred revolutions of the
soliton around the center. Finally, in Fig. \ref{quinta}, it is
shown that if one of the gaussian solitons is replaced by a vortex
soliton the asymmetry of the interaction and the longer
interaction range of the vortex soliton (for which $n\sim
r^2e^{-r^2}$) make this configuration unstable and the initial
configuration is destroyed after a few rounds.

We have also analyzed several soliton molecules in three spatial
dimensions. In Fig. \ref{sixta} we summarize the results for the
situation of a stable configuration made of two weakly interacting
gaussian solitons. The evolution for long times shows that the
inter-component distance $Q(t)$ suffers only small oscillations
[Fig. \ref{sixta}(f)] which manifest on the plots where more
interaction is apparent [Fig. \ref{sixta}(c,d)].

\noindent {\em Conclusions.-} In this paper we have presented
several soliton molecules built up from  scalar solitons of the
trapped NSE. The method presented here allows to generate many
different multidimensional soliton molecules.

Between the possible applications of the vector solitons described
here, in the field of optics they could be used as a way to
transmit different laser beams in a single optical fibers. In any
case the vector solitons presented here represent another step
into the comprehension of complex objects sustained by nonlinear
forces with promising applications in different fields.

 This work has been partially
supported by the Ministerio de Ciencia y Tecnolog\'{\i}a under
grant BFM2000-0521 and Consejer\'{\i}a de Ciencia y
Tecnolog\'{\i}a de la Junta de Comunidades de Castilla-La Mancha
under grant PAC-02-002. V. V. is supported by Ministerio de
Educaci\'on, Cultura y Deporte under grant SAB2000-0256.

\end{document}